# Comment: The Essential Role of Pair Matching

**Jennifer Hill and Marc Scott**

## 1. INTRODUCTION

We appreciate having the opportunity to comment on the well-motivated, highly informative and carefully constructed article by Imai, King and Nall (IKN). There has been a great deal of confusion over the years about the issue of pair-matching, often due to a conflation of the implications of design versus analysis choice. This article sheds light on the debate and offers a set of helpful alternative analysis choices.

Our discussion does not take issue with IKN's provocative assertion that one should pair-match in cluster randomized trials "whenever feasible." Instead we will explore the trade-offs between using the inferential framework advocated by IKN versus fitting fairly standard multilevel models (see, for instance, Gelman and Hill, 2007).

The IKN design-based treatment effect estimators have the advantage of being simple to calculate and having better statistical properties in general than the harmonic mean estimator that IKN view to be the most standard estimator in this setting. Variance estimators for SATE and CATE are not identified, but that is a function of not making the assumption of constant treatment effects, which we find realistic. IKN do provide upper bound variance estimators for these quantities of interest. Perhaps the biggest drawback to these methods is that they are not flexible if it is necessary or helpful to extend the framework to accommodate additional complications or information.

*Jennifer Hill and Marc Scott are Associate Professors of Applied Statistics, Department of Humanities and Social Sciences, Steinhardt School of Culture, Education and Human Development, New York University, 246 Greene St., Room 300, New York, New York 10003, USA e-mail: jennifer.hill@nyu.edu; marc.scott@nyu.edu.*

The strength of multilevel models in this estimation setting is the flexibility to build in complexity that could provide us with additional information, increase our precision, or sometimes even reduce bias (for instance, when correcting for "broken" randomization). As an example, while the IKN variance estimators accommodate varying treatment effects, the multilevel model provides a framework to actually examine these pair-to-pair differences. The model can also be extended to allow treatment effects to vary over covariate-defined subgroups which has the potential to substantially increase our understanding of effect transmission. Conditioning on pre-treatment covariates can also help to increase precision (and even reduce bias in situations where the randomization has been less pristine). Moreover, not only can multilevel models include covariates and random treatment effects quite readily, but the need for such terms can be evaluated statistically.

A further example is the ability of models to accommodate missing data at the individual level (rather than entire clusters being missing due to group-level noncompliance or attrition which IKN address). This can be naturally incorporated into a model-based framework as well; it's unclear how the IKN framework would handle this complication.

Of course, these advantages come at the cost of making some modeling assumptions. IKN go so far as to claim that these approaches "violate the very purpose of experimental work which goes to great lengths and expense to avoid these types of assumptions." However, the primary purpose of experimental work is to avoid the *untestable* assumption of ignorability (or strong ignorability) that is so difficult to avoid in observational work. While it is true that we do not need to build models post-randomization in order to estimate treatment effects, this can hardly be viewed as the goal of randomized experiments. In fact, randomization actually increases robustness to model-misspecification, creating a safer climate within which to build models than would otherwise exist. Moreover, the parametric assumptions we make with a multilevel model are testable, for instance, using graphical regression diagnostics.

It could be argued that multilevel models have the disadvantage of being more complicated to fit. However with the capabilities of current standard statistical software the level of technical expertise required to fit such models is well within the reach of most applied researchers today.

## 2. SIMPLE MULTILEVEL MODELS FOR ESTIMATING TREATMENT EFFECTS

First we lay out a few simple multilevel models for estimating treatment effects in the setting of a pair-matched cluster-randomized experiment. Clearly we have not exhausted all possibilities, but the models we discuss have the advantage of being relatively simple, easily fit with standard software, and readily expandable to more complex settings.

A very simple model for observation $i$ in cluster $j$ and pair $k$ is

$$(1) \qquad Y_{ijk} = \tau T_{jk} + \alpha_k + \varepsilon_{ijk},$$

with a common treatment effect, $\tau$, as well as varying intercepts $\alpha_k$, where $\alpha_k \sim N(\alpha_0, \sigma_\alpha^2)$. $T_{jk}$ is a treatment indicator, and $j \in \{1,2\}$ while $k \in \{1,2,\ldots,K\}$. As is common in multilevel models of this sort, the random terms are assumed independent of the predictors, an assumption which is particularly defensible in the context of a randomized experiment, and $\varepsilon_{ijk} \sim N(0, \sigma_\varepsilon^2)$.

A simple adjustment, allowing $\tau$ to vary by pair, yields a model for heterogeneous treatment effects:

$$(2) \qquad Y_{ijk} = \tau_k T_{jk} + \alpha_k + \varepsilon_{ijk},$$

with $\tau_k \sim N(\tau_0, \sigma_\tau^2)$. We typically do not want to assume that $\alpha_k$ and $\tau_k$ are independent, therefore there is a covariance term in the model, $\sigma_{\alpha\tau}$, and the pair $(\alpha_k, \tau_k)$ are assumed bivariate normal.

A word of caution is warranted with regard to the $\tau_k$. These parameters cannot be interpreted causally except in the special case in which we know that clusters have been perfectly matched on their potential outcomes (which is implausible in practice). Otherwise, we cannot separately identify variation caused by within-pair cluster mismatch from variation that is due to treatment effects that actually vary across pairs. Nonetheless, allowing $\tau$ to vary is important because it allows us to test for this extra source of heterogeneity (whatever the true source of the heterogeneity). To the extent that we can satisfy ourselves that we have indeed obtained close matches (mostly likely after having also conditioned on some highly predictive pre-treatment variables), we can move toward a causal interpretation of these quantities. However, if our goal is to explore treatment effect moderation, we're probably better off doing so by (additionally) allowing the treatment effects to vary by covariate levels.

We can augment either of these models by including cluster-level covariates, $X_j$. This is particularly helpful when we are unable to perfectly match clusters. Here we focus on inclusion of covariates purely for increasing precision (not to explore treatment effect moderation). In this case we add a cluster-specific level to the model, as in

$$(3) \qquad \begin{aligned} Y_{ijk} &= \tau_k T_{jk} + \phi_{jk} + \varepsilon_{ijk}, \\ \phi_{jk} &= X_{jk}\beta + \alpha_k, \end{aligned}$$

where $\phi_{jk}$ captures cluster-specific variation that depends on both $X_{jk}$ and our varying pair intercepts, $\alpha_k$.

## 3. EXAMINING THE IMPLICATIONS OF IMPERFECT MATCHING AND TREATMENT EFFECT HETEROGENEITY

We explore the implications of imperfect matching and the presence of treatment effect heterogeneity through a small set of simulations. Our primary simulations vary the following components: (i) cluster size perfectly or imperfectly matched, (ii) cluster-specific SATE perfectly or imperfectly matched and (iii) treatment effect fixed or varying. Simulations are repeated 100 times for each scenario.

The data generated in each simulation are fit using the two multilevel models laid out in equations (1) and (2) above (we'll refer to them as MLM1 and MLM2, for the constant and varying treatment effect models, respectively). To represent an analysis option that would be easy to use by an applied researcher we fit the multilevel models using the `lmer` command (package is `lme4`) in R (R Development Core Team, 2008; very similar packages exist in Stata, SAS and SPSS, among others) and used the standard estimates. In theory, however, one could fit these models using a more flexible package such as BUGS or JAGS in which case it would be trivial to reweight the $\tau_k$ in order to make inferences about any of a wide range of different quantities of interest. For comparison purposes we fit the IKN SATE estimator (to mirror the multilevel model's implicit weighting scheme by pair sample size) using the upper bound variance estimate to demonstrate the relationship between this bound and the uncertainty estimate in MLM2. Since we can compare nested multilevel models using likelihood ratio tests (LRTs), we also evaluate whether the model detects evidence of variation in treatment effects. We then extend the simulations to incorporate a cluster-level covariate as described in more detail below, and fit the multilevel model described in equation (3) above.

We simulate matched-pair cluster randomized experiments in a manner similar to the IKN simulations with the notable difference that we do not force cluster-specific SATE to be perfectly matched within pair (as described below). The number of pairs in each simulation is 30. Throughout our simulations, the average (or fixed) cluster size is 50. When cluster size is imperfectly matched across clusters, we allow it to vary based on multinomial draws from cluster labels, each equally likely, drawing a sufficient sample so that the expected size of any cluster is 50. Using this strategy, the average difference in cluster size is about 8 and the average standard deviation of these differences across repeated simulations is approximately 6.

Potential outcomes were simulated such that for a given pair $k$, with cluster $j=1$ as the control, and cluster $j=2$ as the treated,[1]

(4) $\quad Y_{.1k}(0) \sim N(\mu_0, \sigma_0^2),$

(5) $\quad Y_{.2k}(0) = Y_{.1k}(0) + \delta_k \quad \text{with } \delta_k \sim N(0, \pi^2 \sigma_0^2).$

---

[1]After submitting a draft of this comment, IKN asked for our code and upon reviewing confirmed an error in our original simulation setup; randomization had not been imposed. Given the necessary restrictions on iterative revisions in this discussion setting we will not update our comment to incorporate the corrected results; however we were permitted to change the description of the simulations to reflect what was actually run (that is, with the error included). We have, however, verified that after correcting our simulations (by imposing randomization such that both potential outcomes and our covariate are independent of the treatment, the situation described in the rest of the discussion), our original conclusions remain; the results are extremely similar to those presented here. More details and the code appear in our online appendix at https://files.nyu.edu/jlh17/public/stat.sci.appendices/.

The error did spark an interesting additional discussion about the potential problems with adjusting for covariates when randomization has failed which IKN explore in their rejoinder.

Therefore $\delta_k$ serves the role of creating imbalance in SATE across clusters (IKN do not allow for this in their simulations). As $\pi$ grows we move from a situation with perfect balance to a situation in which we may as well have randomly chosen pair matches.

Treatment effects were either kept constant across pairs at $\tau_{jk} = 3.2$ for all $j$ and $k$ or were allowed to vary. Heterogeneous treatment effects were generated using a nonlinear deterministic function of the cluster potential outcome under treatment such that $\tau_{jk} = 30/Y_{.jk}(0)$. This creates a partial ceiling effect in which larger baseline values are associated with smaller treatment effects and as such the distribution for both $Y(1)$ and $\tau_{jk}$ are quite skewed (again mimicking the IKN example).[2] The mean of $\tau_{jk}$ across $j$ and $k$ is about 3.2 on average under this formulation.

Individual-level observations are generated from these cluster potential outcomes by adding random errors,

(6) $\quad Y_{ijk}(\cdot) = Y_{.jk}(\cdot) + \epsilon_{ijk},$

with $\epsilon_{ijk} \sim N(0, \sigma_\epsilon^2)$. We chose $\mu_0 = 10$, $\sigma_\epsilon^2 = 1$ and $\sigma_0^2 = 4$ for all simulations.

## 4. SIMULATION RESULTS

In Figure 1, we plot the standard error associated with our three estimates of the common treatment effect when cluster size is not perfectly matched (the scenario in which cluster sizes are equal is nearly identical, with minor differences noted below). Panel A displays the results in the scenario when treatment effects are constant (ignore the thick grey line at this point in the discussion). When $\pi = 0$, match quality is perfect, and multilevel model estimators have the same precision. However, as the match quality degrades (as represented by increasing levels of $\pi$) the lines diverge rapidly with MLM2 reflecting increasingly higher levels of uncertainty.

We might think that MLM1 is the "correct" model in this simulation scenario—after all, the treatment is constant. However, in terms of heterogeneity, there is no identifiable difference between poor matches and variable treatment effects. So in the likely realm of imperfect matches, which model do we prefer, and why? Model selection techniques such as LRTs will guide us toward models that capture variation, when it is present, and we find that when $\pi > 0.13$, the null that $\sigma_\tau^2 = 0$ is rejected at the 0.05 level. We represent this shift away from MLM1 by slowly greying out its standard error in panel A. Thus models can provide evidence of either imperfect matching or variable treatment effects, but with this design, cannot adjudicate between the two. Of course, to the extent that we can use covariates to sufficiently improve across-cluster equivalence within pairs (that is, to make up for imbalance remaining after matching), we might have more confidence in using such tests to infer that the treatment effect is constant.

---

[2]We were able to obtain data from IKN on the distributon of pair specific differences in means in the data they used as a starting point for their simulation. We satisfied ourselves that the distribution in our simulations of the same quantity is even more skewed, thus clearly violates the assumption of normality of the treatment effects built in to our multilevel model.

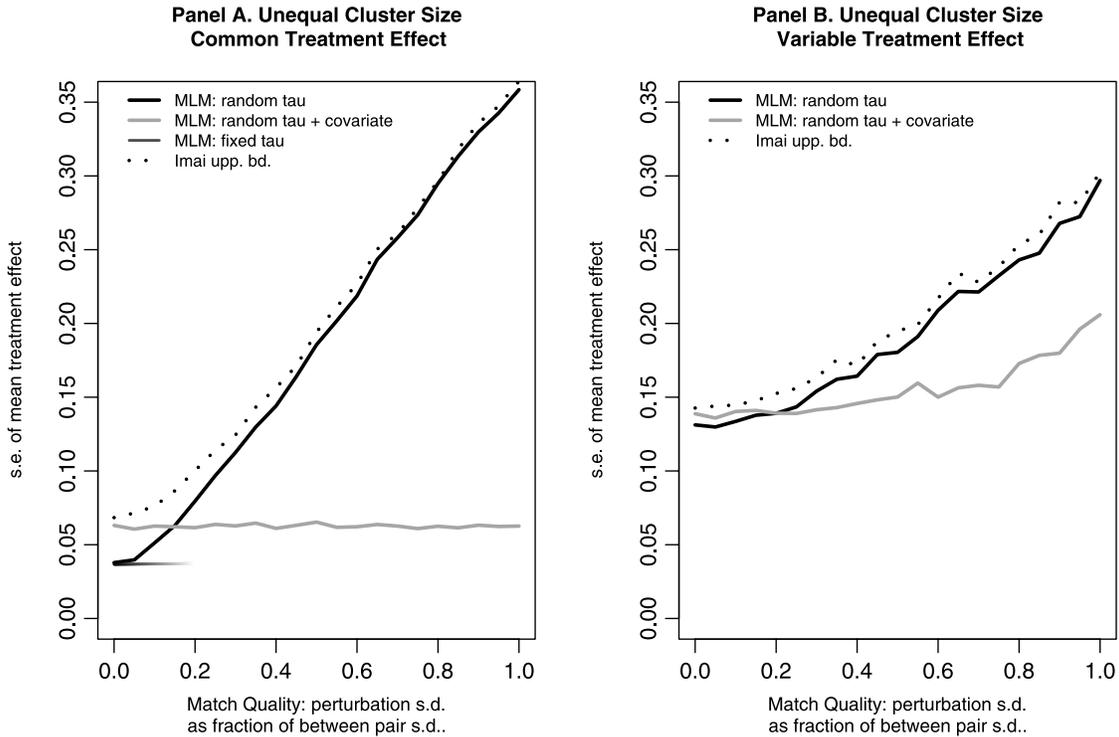

FIG. 1. *Plots that display how the standard error for each method varies with increasing disparity in the matches, as measured by $\pi$. The left panel displays results from the scenario with constant treatment effects; the right panel displays results from the scenario with treatment effect heterogeneity.*

What's interesting to note is that the IKN upper bound variance estimator for SATE closely mimics the uncertainty estimate from MLM2. When cluster size is equal, the IKN estimator's precision is nearly coincident with that of MLM2 (not shown). Of course, neither the IKN estimator nor MLM2 can distinguish between true treatment effects and pair mismatches.

In Figure 1, Panel B, we plot the standard error associated with two of our three estimates of the common treatment effect when cluster size is not perfectly matched and treatment effects vary nonlinearly as specified above. Again, ignore the thick grey line at this point in the discussion. When $\pi = 0$, there is already a difference in precision between MLM1 and MLM2. MLM1 completely ignores any heterogeneity, so with this incorrect assumption, it underestimates the uncertainty. Given that LRTs would correctly suggest that MLM1 is insufficient, in other words, $\sigma_\tau^2 > 0$, we do not include MLM1's precision in the plot.

We now concentrate on MLM2 and the IKN estimator, and again we see that the precision follows a comparable trend as $\pi$ is increased and matches degrade. Imperfect matches and the varying treatment effects are increasingly confounded, and the uncertainty concomitantly increases. It is somewhat surprising that the precision curves degrade at a slower rate than those in Panel A, yielding superior precision when $\pi > 0.5$. This can be attributed, however, to the additional information contained in the correlation between treatment and pair effects, created by the nonlinear transformation that generates $\tau_{jk}$. We confirmed this using a different simulation setup for which variable treatment and pair effects were generated independently (not shown); in these simulations, the precision degrades a bit more quickly than in the case of common treatment effect, as expected. That MLM2 and the IKN estimator's s.e. are at higher levels in Panel B when $\pi < 0.5$ is simply the effect of increased baseline variation in treatment effects introduced in the simulations with variable $\tau_{jk}$.

An equally important point here is that even though we in essence incorrectly model the skewed treatment effects by pretending they are normally distributed, this "model failure" did not introduce bias or reduce precision (skew may have induced slight overestimation of the variance in the treatment).

### 4.1 Covariates

We operationalize covariates as having partial information on $Y_{\cdot jk}(0)$, but not directly on $\tau_{jk}$. The simplest way to do this, in our simulations, is to set $X_{jk} = Y_{\cdot jk}(0) + \zeta_j$, where $\zeta_j \sim N(0, \sigma_\zeta^2)$ is a noise process that limits our ability to recover the level of the potential outcome. When $\sigma_\zeta^2$ is small, we should eliminate, or nearly eliminate the variation between pairs, $\sigma_\alpha^2$. This should result in increased precision for the treatment effect, particularly when treatment and within pair differences are greatly confounded. In the simulation results shown in Figure 1, we chose $\sigma_\zeta^2 = 0.2^2$, which is large enough to obscure some information in the covariate, but not so large as to render it nonsignificant, and fit the model given in (3) above. The standard errors for treatment effects (our primary assessment) are presented as a grey line which is remarkably constant across various levels of match quality. Panels A and B are quite similar, so our remarks apply to either. When match quality is very good, conditioning on covariates actually adds a small amount of uncertainty to the treatment estimate. However, the payoffs associated with covariates include: dramatic reduction of between pair variance $\sigma_\alpha^2$, which provides the opportunity to identify a simpler (common treatment) model, when this actually is the case, and improved precision when match quality is poor. To summarize, the impact of a (significant) covariate or set of covariates should be to decrease the variance $\sigma_\alpha^2$, and this has the potential to yield remarkable precision gains.

## 5. CONCLUSION

In some ways, the IKN framework is actually quite similar to the multilevel framework that allows for variation in treatment effects across pairs. The advantage of the multilevel framework however is in moving beyond the simplest scenario to incorporate additional complexity for greater precision or greater understanding. We have illustrated only a small number of the potential set of such model expansions.